\documentclass{article}

\usepackage{PRIMEarxiv}

\usepackage[utf8]{inputenc} 
\usepackage[T1]{fontenc}    
\usepackage{hyperref}       
\usepackage{url}            
\usepackage{booktabs}       
\usepackage{amsfonts}       
\usepackage{nicefrac}       
\usepackage{microtype}      
\usepackage{lipsum}
\usepackage{fancyhdr}       
\usepackage{graphicx}       
\graphicspath{{media/}}     

\title{Theme and Topic: How Qualitative Research and Topic Modeling Can Be Brought Together}

\author{
  Marco Gillies \\
  Department of Computing \\
  Goldsmiths, University of London, UK \\
  \texttt{m.gillies@gold.ac.uk} \\
   \And
  Dhiraj Murthy \\
  School of Journalism and Media \\
  University of Texas at Austin, USA\\
  \texttt{dhiraj.murthy@austin.utexas.edu} \\
  \And
  Harry Brenton\\
  BespokeVR\\
  London, UK \\
  \texttt{harry@bespokeVR.com} \\
  \And
  Rapheal Olaniyan \\
  Department of Computing \\
  Goldsmiths, University of London, UK\\
  \texttt{rolan001@gold.ac.uk} \\
}

\begin{document}
\maketitle

\begin{abstract}
Qualitative research is an approach to understanding social phenomenon based around human interpretation of data, particularly text. Probabilistic topic modelling is a machine learning approach that is also based around the analysis of text and often is used to in order to understand social phenomena. Both of these approaches aim to extract important themes or topics in a textual corpus and therefore we may see them as analogous to each other. However there are also considerable differences in how the two approaches function. One is a highly human interpretive process, the other is automated and statistical. In this paper we use this analogy as the basis for our Theme and Topic system, a tool for qualitative researchers to conduct textual research that integrates topic modelling into an accessible interface. This is an example of a more general approach to the design of interactive machine learning systems in which existing human professional processes can be used as the model for processes involving machine learning. This has the particular benefit of providing a familiar approach to existing professionals, that may can make machine learning seem less alien and easier to learn. Our design approach has two elements. We first investigate the steps professionals go through when performing tasks and design a workflow for Theme and Topic that integrates machine learning. We then designed interfaces for topic modelling in which familiar concepts from qualitative research are mapped onto machine learning concepts. This makes these the machine learning concepts more familiar and easier to learn for qualitative researchers. 
\end{abstract}


\section{Introduction}

This paper investigates an analogy between two seemingly different research methods: \emph{qualitative research} and \emph{probabilistic topic modeling}. Qualitative research involves detailed reading of texts resulting in a rich human interpretation. It can give highly nuanced and insightful analyses but is very time consuming and generally cannot be scaled up to the volumes involved with ``Big Data''. Topic modeling on the other hand is an automated method that is very well suited to big data, but which lacks the nuance of human interpretation. However, we argue that they both share a common goal: discovering a number of underlying themes within data. This paper investigates this analogy and ask whether it is possible to use this analogy to bring the two methods closer together. We assess the possibility critically and raise as many questions as it answers, while being a starting point for future research. 

This analogy presents an example of a more general approach to designing interactive systems that are based on machine learning by taking an analogy between a machine learning approach and an existing human task. Following in this vein, we seek to design software that is more accessible to existing professionals than a traditional machine learning system, which can be daunting at first. In particular by working around existing concepts and workflows we may present the elements of a machine learning system to users in a way that is readily interpretable using their prior knowledge.  

This paper presents an example of this design approach via the design of a research system that integrates machine learning with workflows drawn from qualitative research. One pitfall of this approach is that inevitably the machine learning algorithm and the human task will differ in potentially subtle ways that could result in confusion or be misleading. Another important challenge of this approach is to be aware of these potential differences and just as we highlight similarities in the design of an interactive interface we must particularly highlight cases where the actions of the machine learning algorithm may differ from what will be a standard human interpretation. The example presented in this paper highlights both these factors of similarities and differences.  

\section{Qualitative Research}

Qualitative Research is a name given to a wide range of research techniques in the social sciences and related disciplines (including HCI\cite{Furniss2011}) based on a detailed, human reading of textual or similar data with the aim of developing themes or theories that take a qualitative form. Qualitative methods are often contrasted with quantitative research methods based on statistical analysis\cite{Mahoney2006} (which generally take the form of hypothesis testing, and so differ from the machine learning techniques often used in big data analysis). Qualitative research encompasses a very wide range of diverse methods and methodologies \cite{Glaser1967,Braun2006,Heath2010,Smith2015}. In this paper we will focus on two of the most popular and broadly applicable: Grounded Theory\cite{Glaser1967,Strauss2008,Charmaz14} and Thematic Analysis\cite{Braun2006}.

Grounded Theory (\cite{Glaser1967,Strauss2008,Charmaz14}) is a methodology for developing fully formed, novel theories that are ``grounded'' in a close analysis of qualitative data. It has been used extensively in HCI\cite{Furniss2011}. Grounded Theory is usually an emergent process. Specifically, data collection and analysis can involve several passes, new variables, and emergent research questions\cite{Murthy2016}. Many diverse approaches to grounded theory have emerged over the years, for example the divergent approaches of the two founders Glaser\cite{Glaser1998} and Strauss\cite{Strauss2008} or the constructivist approach of Charmaz\cite{Charmaz14}. These all differ as much in their epistemological foundations as they do in their practical methods. In this paper we will attempt to focus on the commonalities between approaches. Thematic Analysis\cite{Braun2006} on the other hand aims not for a full theory but and understanding of ``themes'' that emerge from the data, which are generally at a lower level of analysis than a full theory. Thematic analysis may variously be thought of as a stage in full grounded theory that precedes full theory development, a form of grounded theory ``lite'' that does not go all the way to theory development, or, as Braun and Clarke\cite{Braun2006} a method in its own right, with a different set of aims. We begin our discussion with Thematic Analysis, because it identifies a number of methods that a common across many qualitative research methodologies. Second, we believe the analogy with Topic Modeling is closer as the themes of thematic analysis are at a similar conceptual level to the topics of topic modeling. Developing full qualitative theories is well beyond the scope of current machine learning. we will then discuss important differences with full grounded theory. 

Grounded theory and thematic analysis both begin with a period of \emph{familiarization} with the data in which researchers begin by reading the data as a whole to get an overall sense of what is being said before doing detailed analysis. This if following by a process of coding which involves a close reading of the data. The research selects important passages in the data and applies ``codes'' to them. Codes are single words or short phrases that summarize and identify the topic of the text. Codes should be sufficiently general that they can apply to multiple parts of the text and so can bring together different passages that are about the same thing. Once an initial close coding has been performed, the researcher goes back through the codes in an attempt to find higher level themes, by combining codes and looking at their relationships. This stage includes a number of variants and different terminologies, Strauss and Corbin\cite{Strauss2008} refer to finding ``concepts'', Braun and Clarke\cite{Braun2006} to searching for ``themes'' and Charmaz\cite{Charmaz14} to ``focused coding'' (in the rest of this paper we will refer to ``themes'' on the understanding that these can stand for a number of concepts with diverse philosophical foundations). Once a number of themes have been discovered, they are reviewed and refined by going back to the original data and comparing to see how well it matches the data. 

In thematic analysis the aim is to produce a number of refined themes. However, grounded theory aims to go deeper and develop theories that relate and explain the themes. This can use a number of further approaches such axial\cite{Strauss2008} or theoretical coding\cite{Charmaz14}, but the most important method is \emph{Constant Comparison}: passages of data are compared to other passages, codes are compared to data, code to other codes and themes to both codes and data. The aim of this comparison is to understand relationships between themes and their relationship to data in order to deepen the researchers' understanding of the phenomena being studied. This is a detailed process of interacting with the data, reading, theorizing and comparing to refine the theory. Another important part of the Grounded Theory process is \emph{theoretical sampling}: data is not collected prior to analysis but as part of an iterative research process in which initial analysis informs the questions and approaches used in later data collection. Later data is collected specifically to better understand the themes discovered in earlier analysis. 

Qualitative research can provide a very rich, nuanced and human understanding of complex phenomena \cite{robertson2019using}. It can highlight subtle and particular themes that can be lost in statistical analysis and can be open to surprising phenomena in a way that hypothesis testing cannot. However, it also has problems. It is an extremely labor intensive process requiring an expert to do very close reading(s) of the data. The time taken limits the scope of what is possible with qualitative research, making it unfeasible for even medium sized data sets, let alone the big data setting where even an overview reading of the entire dataset is not possible \cite{murthy2015critically}. 

\section{Topic Modelling}

The problem of data size means that big data analysis is done by  machine-based approaches. One of the most popular is \emph{Probabilistic Topic Modeling}\cite{Blei2010a}, a set of machine learning approaches that analyze large corpora of textual data, consisting of many individual documents, to extract the underlying topics or themes within the data. Topic modeling has been applied to a number of domains such as the analysis of academic literature\cite{Blei2010a}, social media ``big data''\cite{xin2019assessing}, news stories\cite{Hoque2015} or transcripts of crisis counseling sessions\cite{Dinakar2015}. 

One of the most popular methods for topic modelling is Latent Dirichlet Allocation (LDA)\cite{Blei2003}. That being said, much newer, deep approaches such as BERTopic \cite{grootendorst2022bertopic} are quickly growing in popularity, particularly in social media applications. Unlike many earlier machine learning methods, LDA allows individual documents to contain multiple topics, and not be about just one thing. A full description of the algorithm is found in Blei \emph{et al.}\cite{Blei2003} here we will highlight a number of important features. In LDA, topics are represented as probability distributions on words, for example the word "education" may be much more likely to appear in one topic than another. Documents are represented as mixtures of topics, with the probabilities of words being determined by the probability of that word in a topic multiplied by the probability of the topic within the document. LDA models are learned using an Expectation Maximization algorithm, which alternates a phase of determining the probabilities of words within a topic (essentially a process of counting words) based on the assignment of documents to topics with a phase of reassigning the documents to topics based on their word probabilities. 

As mentioned in the introduction, there is an interesting analogy between the aims of qualitative research and topic modeling. In his survey paper, Blei\cite{Blei2010a} describes topic models as: \emph{``statistical methods that analyze the words of the original texts to discover the themes that run through them, how those themes are connected to each other, and how they change over time.''}. Not only is the word ``themes'' directly analogous to Braun and Clarke\cite{Braun2006} but the focus on themes, the relationships and variations is very close to qualitative research ideas of constant comparison between themes and data. 

However, there are many differences. The automated nature of topic models make them applicable to very big data, but in many ways it also means that it is  impoverished relative to the rich human interpretation involved in qualitative research. For example, LDA essentially consists in counting words and misses much of the contextual understanding that human reading gives, it does not even take account of the order of words in sentences, the only contextual information used is the co-occurrence of words in documents. 
\section{Interactive Machine Learning}

If machine learning methods like topic models can handle very large data sets but ignore the complexities available to human interpretation within qualitative research, is it possible to bring the two methods closer to bring the benefits to both (after all isn't the aim of HCI to bring humans and computers closer)? Recently the challenge of reconciling human and machine learning processes has been studied in \emph{Interactive Machine Learning}\cite{Fails03,Amershi2014} and Human-Centred Machine Learning\cite{Gillies2016}. 

Machine Learning is traditionally viewed as a batch process in which a large, pre-existing dataset is fed into the algorithm, which processes it and returns an answer. The role of the human in the process is simple to collect the data, ideally in as passive a way as possible to ensure that the data is independent and identically distributed. Interactive machine learning on the other hand sees the role of the human as much more active. They actively select data items to be most representative and appropriate to the task. The selection of data is not done prior to running the algorithm, but in a tight interactive loop with machine learning. The human selects a small amount of initial data which the computer uses to generate an initial model. The human then adds more data specifically to refine the model and correct errors in it. This new data is not arbitrary or randomly selected, but is specifically chosen, through human judgment, as the best data to improve misconceptions in the learned model (this may seem similar to active learning, but is fundamentally different because it is a human that judges the usefulness of a data item, not a computer). In fact, the human may not know ahead of time exactly what they want the computer to learn, their concept of the problem also develops via interaction with the learned model and finding new data. 

This interactive process is much closer to qualitative research and particularly grounded theory. Human judgment is vital to both. Both involve an interactive process of constant comparison. In qualitative research, it is checking themes against data. In interactive machine learning, the model is tested against new data. Data collection is not prior to analysis but is done interactively, with new data being sought based on the outcomes of analysis. 

Human judgment is also important in both machine learning and topic modeling. However this human aspect of the process is often underplayed and  somewhat hidden in academic publications that focus on the success of the algorithms and those which measure that success on standardized datasets, without explaining what human intervention was involved in the training. In topic modeling, in particular, the human interpretation tends to be focused on the final stage: the interpretation of the output of the algorithm. There is likely also to be considerable interpretation in the process of training and debugging topic models. However, the human side of that process has been  minimally studied. An interactive human-centered approach to machine learning allows us to foreground that human intervention and in particular look at ways in which using existing models of human interpretation, such as qualitative research, can help us build better systems that make it easier and more productive to integrate human interpretation into the process of topic modeling.

This suggests that applying interactive machine learning approaches to machine learning could be valuable. In fact, there has been some existing work on interactive topic modeling. For example, Hoque and Carenini\cite{Hoque2015} allows users to interact with a visualization of a topic model of new items; Hu \emph{et al.}\cite{Hu2011} allow user to add constraints to a topic model in order to influence the learned model; and Dinakar \emph{et al.}\cite{Dinakar2015} use a mixed initiative approach which interleaves unsupervised topic modeling with human labeling of data. Chen \emph{et al.} have also explored the differences between qualitative research and machine learning, and particularly  regarding the role of ambiguity\cite{Chen2018}.

\section{Themes and Topics: an analogy}

 Ultimately, thematic analysis and topic Modeling have remarkably similar aims: to identify a number of underlying themes in (normally) textual data. However, they differ greatly in how they achieve this. Thematic analysis is a highly interactive process involving rich human interpretation, while topic modeling is an entirely automated, algorithmic method. However, interactive machine learning approaches aim to increase the role of human interpretation and interactive development of models within machine learning. This raises the question of whether it is possible to use interactive machine learning to create analytic methods that combine some of the benefits of both qualitative research and topic modeling. Can topic modeling algorithms be influenced by human judgment? Can we develop methods that are based on human judgment but are applicable to very large datasets? 

These questions lead us to an approach to the design of machine learning systems in which we use an analogy to a human process  (in this case qualitative research) in order to design a machine learning process that is more readily accessible to professionals in an existing domain  (in this case social science research). This approach has two main strands: workflows and interfaces.  

First, we investigate the workflows required to perform  various tasks  related to undertaking qualitative research and how that relates to the workflows of machine learning. We  then design new workflows that are related to the existing workflows of qualitative research, but which integrate  machine-based methods. 

The second major strand is the design of actual interfaces that professionals will interact with as they perform these workflows. These interfaces must support conceptual models that are related to existing conceptual models\cite{Norman13,Blandford08} in the domain of qualitative research.  In other words, the concepts presented in the interface must relate closely  to concepts familiar to qualitative researchers, while also representing elements that actually exist in machine learning algorithms. In  our case, we  have to represent concepts (e.g., machine-rendered topics) in ways  accessible to qualitative researchers (e.g., codes and themes). This approach is inspired by Blandford's CASSM framework\cite{Blandford08}, which stresses the need to match the concepts used by humans to the concepts present in the interfaces and underlying implementation of the software that they use. 

The next section  provides an outline of possible workflows that combine qualitative research and topic modeling before describing the design of a prototype that addresses some of the requirements of the workflows. The following section describes the design of an interface to support these workflows

\section{Possible Workflows}

Having described a possible integration of topic modeling and qualitative methods,  it is important to examine  practical applications. This section discusses four possible workflows for the technologies described. The workflows differ primarily in terms of what the desired output looks like: does it look like a number of qualitative themes (which were developed with the help of topic modeling) or a number of quantified topics (which are informed by qualitative research). 
Each workflow has its own requirement that will influence the design of a prototype. 

\subsection{Topic Modelling as a seed for Qualitative Research}

A first possible workflow is to use a topic model as a starting point for qualitative research. A researcher could look at the topics discovered to give an initial understanding of the data. This would fit well in the initial familiarization phase of the research and enables researchers to gain a familiarity with a dataset that is too large to read. Researchers could explore a random sample of topics. Also,  they could discuss topics with respondents and  evaluate whether they can discern any meaning in the topics.  This could potentially empower respondents, making them  stakeholders  who have a role in evolving the research questions. 

The main requirement of this workflow is the ability to generate topics from data and then explore those topics. This exploration requires the ability to find documents that are associated with a topic and to understand why they were associated with that topic. Since topics that are generated without human input will not have names or any prior semantics, this exploration will largely be a process of understanding what each topic means. In many ways, this is similar to the process of familiarization common in qualitative research,  which includes finding and  interpreting documents associated with the topic to understand them. The major difference  is a focus on the reasons that documents are assigned to topics (since each document can contain several topics),  rather than simply on the document.  Researchers therefore  must be able to see which words were primarily responsible  for the assignment. 

\subsection{Qualitative Research as a seed for Topic Modeling}

The problem of initial conditions in topic modeling suggests another possible workflow. A qualitative analysis of a subset of the data could be used as an initial seed for running a topic model. This is likely to improve the quality of the topics found by grounding them in a human understanding of the data. Some degree of human judgment does seem to be used already in topic modeling, for example Hall, Jurafsky and Manning\cite{Hall2008} say they \emph{``hand-selected seed words for 10 more topics to improve coverage of the field''} and Dinakar \emph{et al.}\cite{Dinakar2015} used expert judgments in their process. The use of a method modeled on qualitative research would encourage a more rigorous and detailed process than is currently used and also enable input from qualitative researchers who are likely to have more experience work than machine learning researchers. 

This workflow is in many ways similar standard qualitative research. After an initial process of familiarization a researcher will code the data and gradually develop these codes into more comprehensive themes. This requires an interface for coding textual data and working with low level codes to develop them into themes. The major difference is that these codes are not simply for human consumption but also serve as input to a topic model. This involves a correspondence between themes and topics. As we have discussed, themes and topics are similar  and this correspondence can be straightforward  (i.e., a topic is created for each theme). There are several caveats to this. First, it requires an algorithm that is able to use labeled data as a starting point for generating topics (see section \ref{sec:implementation} for more detail). Second,  though in qualitative research themes are often defined in terms of low level code, the topic models must be defined in terms of the underlying data. Finally, as in all of these workflows, it is important to bear in mind interpretation and topic modeling algorithms.  in particular, the topic model may not be able to  identify the meanings that the human researcher used to define certain themes.

\subsection{Interactive Topic Modeling}

While there may be value in using one method as a starting point for the other, the real benefit is likely to come if we enable a tight interaction between the two. Machine learning algorithms often work very differently from human interpretation and that results are commonly not what was expected. While the words ``someone'' and ``stops'' may occur frequently in documents relating to ``dating'' a human researcher would likely not judge them as words that identify the topic ``dating''. A single pass of the algorithm would commonly throw up such problems and it would be very difficult to identify and correct these problems in advance. An interactive approach would make it possible to detect and correct problems as they occur:  building up a topic model through interleaved steps of coding the data and training a topic model on the new codes. Errors in the topic model can be corrected by supplying new codes and the researcher could have much greater agency and control over the technology than would be possible in a batch process. The interactive element also makes it possible to take account of how researchers' understanding of the theme evolves over time. Seeing the computer's interpretation of the theme can help to better understand what is and is not part of that theme and what that theme means. This is the positive benefit of the unexpected nature of the machine learning results. Moreover, whether they are correct or wrong, they can help bring greater insight by challenging a researcher's initial understanding. 

This workflow has many similarities to the previous two. It involves coding and using those codes as input to a topic model. It also involves exploring and understanding the resulting topic model. However, it also involves changing that topic model. This requires an algorithm that update topic models based on new input (while keeping them otherwise as similar as possible). It also requires mechanisms for editing. The most basic mechanism is to code more data and therefore provide more data as input. This can help address situations where the topic model fails to label a document with a certain topic. However, it must also be possible to correct situations where the topic model labels a document with an incorrect topic. This means adding negative examples to a dataset, which is less common in machine learning.


\subsection{Topic Modeling as a support for Qualitative Research}

 Though the previous workflow aims to create a topic model, that is not the only possible output. The topic model could simply be used as a support for a more purely qualitative research process. For example, a topic model could gradually be learned based on the researchers' codes and this could be used to make suggestions for new documents via a more sophisticated search function than simply using keywords. This would have the benefit of allowing researchers to better navigate very large datasets and do more effective theoretical sampling. 
Again, the value of this approach is highly reliant on the quality of the topic model. A poor model is likely to give suggestions that distract or even harm the research process. This is likely to be particularly problematic in the early stages of research  (e.g.,  ideation) when there is little coded data and the topics are likely to be low quality. A basic literacy of topic modeling and its limits amongst a research team  does help ameliorate this.

This workflow shares many of the requirements of the previous one, but is likely to be more implicit. Researchers are less likely to want to explicitly correct the topic models, so the algorithm will be more autonomous. A key challenge, at least in early stages,  is that the suggestions are likely to be irrelevant.  Therefore, a design  sensitive to maximizing relevancy should be used to avoid overloading researchers with a bundle of distracting topics. 

\section{Theme/Topic: an interactive prototype}

\begin{figure}
  \centering
  \includegraphics[width=\linewidth]{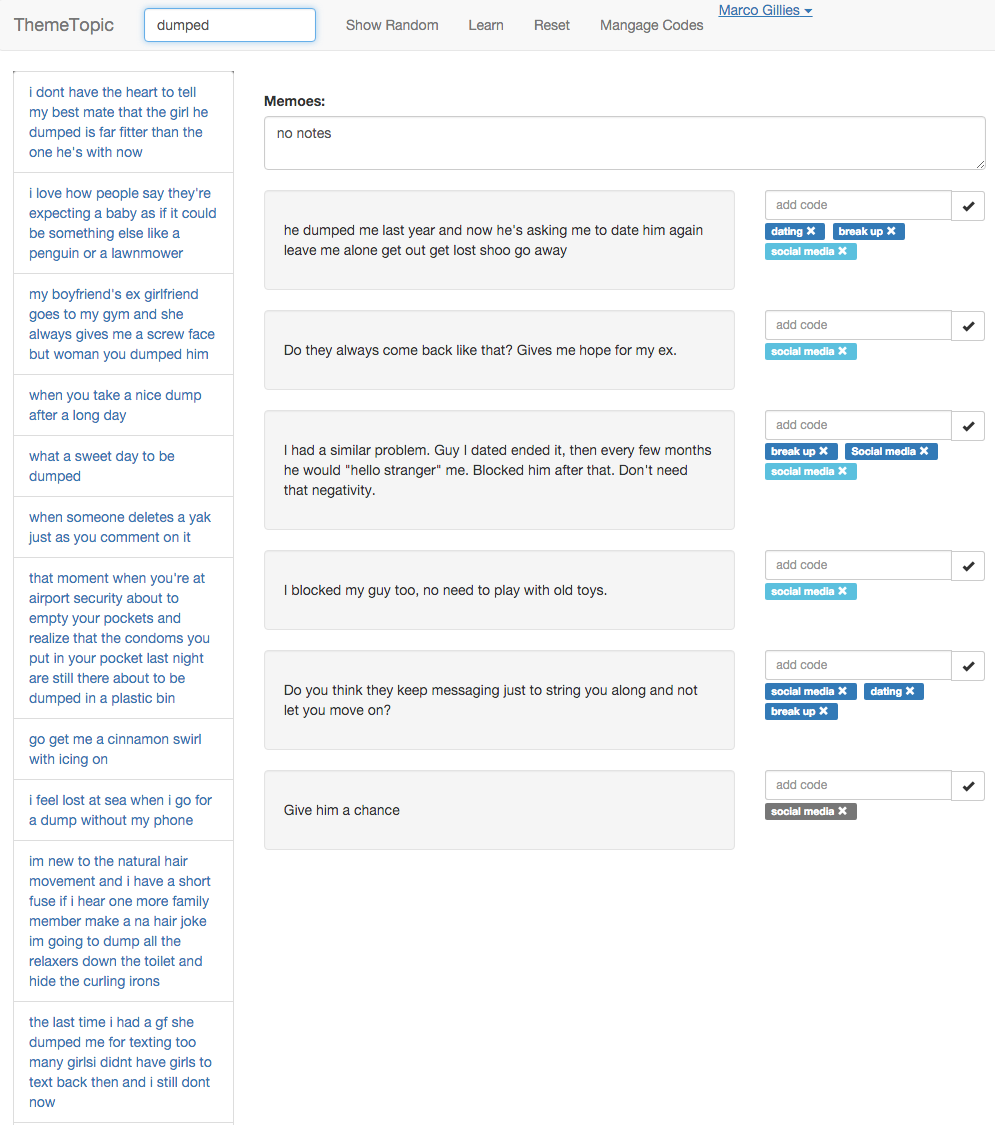}
  \caption{The main theme/topic interface.}
  \label{fig:interface}
\end{figure}

To better understand the possibilities of combining topic modeling and qualitative research, we have developed the  theme and topic prototype research software  depicted in figure \ref{fig:interface}. The interface is  optimized for the analysis of social media data, which consists of short pieces of text arranged into conversation or comment threads. 

\begin{figure}
  \begin{center}
  \includegraphics[width=0.6\linewidth]{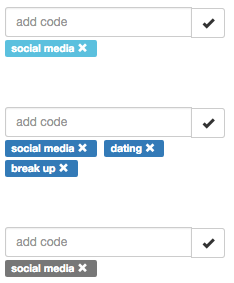}
  \caption{Codes added to the side of a document. The dark blue codes have been added manually, the light blue codes have been added automatically and the gray topics have been deleted manually.}
  \label{fig:codes}
  \end{center}  
\end{figure}


Our explicit aim was to design a machine learning system around a human process. Rather than focusing on a traditional machine workflow, we designed  our theme and topic software around the process of qualitative research and the first three workflows described in the previous section (the fourth one being rather different in character). We  structured the interface around the concept of a ``code'' which is central to  grounded theory-based qualitative research. Coding is the most fundamental part of  this research process, and the primary means of understanding data. All higher level concepts such as themes and theories  tend to be derived from codes (indeed, themes can be considered a form of code in axial coding).  The interface is structured around the various stages of  this type of qualitative research process,  which we describe in this section. 


\subsection{Initial Familiarization:}  As shown in figure \ref{fig:interface},  the data to be interpreted is displayed. The center portion displays the text of a document. In our use case, a social media conversation  is divided into individual comments. The corpus of documents can be explored via a list on the left hand side. As the dataset is large only a small proportion can be displayed at a time. It is also possible to search the whole corpus based on keywords or other  attributes such as geolocation. 

\begin{figure}[t]
  \begin{center}
  \includegraphics[width=\linewidth]{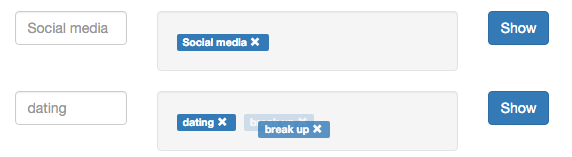}
  \caption{The interface to managing codes and themes. Codes can be added to themes to create higher level, merged themes. The code ``break up'' is being dragged onto the theme ``dating'' to add it.}
  \label{fig:managingcodes}
  \end{center}  
\end{figure}

\subsection{Coding:} Users are able to select passages of the text and apply codes to them. The codes take the form of individual words or short phrases. They appear in dark blue in a box on the right hand side of the text (figure \ref{fig:codes}). Whenever a new code is added (i.e. a word or phrase that has not previously been used as a code) a new topic is created for it in the topic model. As the topic models learns, it  applies topics to portions of the text and these topics will also appear in the codes box as the name of the topic, in light blue to distinguish them from the manually added codes. Finally, users may also delete the automatically added codes to indicate that they  do not apply to the text  being coded. These deleted codes appear in gray. Clicking on any of these codes will highlight all words in the document that have been associated with that code.

\subsection{Finding themes:} Once the low level codes have been applied, they can be reviewed and themes developed. Figure \ref{fig:managingcodes}  illustrates our interface for managing codes and themes. Each line represents a theme with a name and a number of codes. Each theme corresponds to one topic in the topic model. Initially there is one theme per code, but codes can be merged into large themes by dragging one code onto another. When this is done the two underlying topics are merged. However, the codes are still represented separately so it is possible to split them again. The names of each theme (and topic) is initially set to the name of the code, but can be changed.

\subsection{Reviewing themes and theoretical sampling:} Themes must be compared back to data. A user can select new texts to  examine and  visualize how the themes apply to them (including the themes that are automatically added by the algorithm). Since social media data is often scraped prior to analysis, theoretical sampling in the traditional sense is often  challenging. However, since researchers can only view a small subset of the data, theoretical sampling can be viewed as a method of deciding which texts to code next. The searching functions allow users to explore the text in theoretically driven way, but the topic model can also support this. The current topic will be the computer's best guess  of which texts contain a given theme. This makes it possible to search specifically for texts that are likely to be relevant to a particular theme. 


\section{Technical Implementation} \label{sec:implementation}

 Our theme topic software's backend of the software uses a variant of Blei's variational LDA algorithm\cite{blei2003latent} to analyze the data into topics. The algorithm we used is semi-supervised in the sense that it can use data labels to guide learning, but does not require all data to be labeled. If none of the data is labeled, then it infers topics using the standard LDA algorithm. If some data are labeled, then these labels influence the learning (see below for more details). The labels are created by a human user via an interface modeled on the grounded theory research process. 

\subsection{Semi-supervised learning}

When a topic is created from a theme, the manual codes influence the learning process in two ways. Firstly, the EM algorithm needs a starting point from which to learn. These initial conditions can have a strong influence in learning. The initial condition for the topics are set to have high probabilities for all of the words that have been coded with that theme and low for all other words. After the first run of the algorithm, the last learned topic model, augmented with any new codes, is used as the initial condition.  This ensures that learning always starts close to the current state. 

The other influence is that the EM algorithm is constrained. The expectation step involves calculating the probability of any given word in a document belonging to a topic. If a word has been coded with the corresponding theme, the probability is set to be 1 rather than calculated using the model (if it is coded with multiple themes, the probability is divided evenly between them). If a code has been explicitly deleted, it must be prevented from appearing in the topic. The probabilities in the initial condition and during the  expectation step are set to 0 to remove all influence. This active element to deleted codes is the reason they are shown in gray in the interface.

\section{Limitations}

We developed theme and topic to interface to a machine learning algorithm, but based on the analogous pre-existing human process of grounded theory, a  qualitative research  method. This was done so as to make the process better suited to qualitative researchers  working with grounded theory approaches. Doing so always risks introducing ``seams'' where the working of the technology is inconsistent with people's conceptual models. The implementation and use of our theme and topic system has highlighted a number of these. 

\begin{figure}[t]
  \begin{center}
  \includegraphics[width=0.7\linewidth]{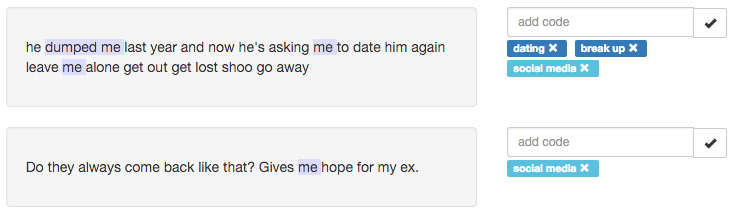}
  \caption{Clicking on a code highlights all of the text associated with that code. This demonstrates that the text is not only what was originally selected, but  includes other uses of the same word. Here, the word ``dumped'' is related to the ``break up'' code, but the word ``me'' is used in ays that are not  explicitly related.}
  \label{fig:highlightedwords}
  \end{center}  
\end{figure}

 In grounded theory, words are treated in context and their meaning is considered unique to that context. On the other hand, the  machine-derived topics of LDA treat all appearances of a word in a document as equivalent ignoring word order. We underscore this by selecting all occurrences of a coded word when a particular code is selected (see figure \ref{fig:highlightedwords}). LDA can handle this to a degree as words can belong to two topics and can represent multiple meanings.  However, users need to be supported in using this feature. A related issue is that LDA represents topics as a probability over \emph{all} words in a corpus. This results in issues described above where the words ``someone'' and ``everything'' are associated with the ``dating'' topic due to common occurrences in dating related documents, but on their own they are not indicative of dating leading to misclassification. This could be a difficult issue for users to grasp. 
 
Learning topics based on a  user's themes is likely to be inaccurate. Initial stages of learning are likely to be poor matches to the  user's concepts. Even late in the process the matches are unlikely to be perfect. It is therefore important to signpost the likely problems and support users in understanding why the algorithm does not work as expected. Displaying the automatically generated codes and allowing  users to see which words they correspond to is a starting point in doing this, as does the ability to interact with the learning process. However, good signposting is likely to be a key challenge for this type of system.

\section{Future Work}

Future work wouuld  therefore be valuable in providing possible solutions to these seams. One suggested by Gillies\cite{Gillies15IUI} is that algorithms should be tailored to interactive contexts and the needs of users, rather than simply putting an interface on an existing algorithm. In this case,  one could investigate algorithms that allow more specification of context. Another approach  would be to allow users to specify more information to the algorithm than simply labels (as Talbot et al. do \cite{Talbot09}).  This could provide the ability to exclude uninformative words like ``someone''. Finally, seams that cannot be eliminated could be made explicit to users through interfaces that support them in understanding the properties of the algorithms and how best to use them.

When deleting a code from a text, it is not enough to just remove it.  Rather, the algorithm must actively prevent the word(s) from appearing in a topic. This could be potentially confusing to people  as often deletion  indicates that something no longer  matters or retains influence. This is why deleted codes are shown in the interface as gray to indicate that they still potentially have influence and meaning. Future work could help refine how deleted codes could be  better integrated into this type of  interface.

\section{Conclusion}

This paper has proposed an analogy between aspects of qualitative research and probabilistic topic modeling. It has developed this analogy into a proposed interface for integrating the two. This approach has the potential to initiate new forms of research that combine the benefits of human interpretation with those of automated processing. However, we also need to consider the problems and challenges, both technical and social, of this approach.







Though integrating qualitative research and topic modeling is  challenging,   interventions can provide valuable contributions to research methods. 
This  type of work is particularly challenging as  it crosses two research methodologies, qualitative and quantitative, which have very different philosophies and  can  sometimes be antithetical. While bringing qualitative and quantitative research is often seen as a holy grail in the silos seen in some social science disciplines, each side has legitimate concerns. Qualitative researchers would worry about mechanization and simplification of complex human phenomena, while quantitative researchers would have concerns about the introduction of subjectivity in the process. These concerns should be taken very seriously in any attempt to combine qualitative research and quantitative big data analysis. A related problem is that any new research method needs to establish its validity and rigor, particularly when it comes to publishing research. This is likely to be a slow process involving a very clear explanation of the methods used. 


An opposite problem is that qualitative researchers may use technology uncritically. New research may be guided too closely by the topic models and not have the literacy of the technology to be critical of an algorithm's output, even though the topic model may be inaccurate or even conflict with their instinctual reading of their data. A key challenge of this type of research is therefore to raise literacy of the research team collectively and to foster critical readings of topic modeling output. All this points to the need for  further work that is sensitive, nuanced,  and is geared towards seeking  meaningful understandings of human research processes and the needs of  qualitative research. 



\section*{Acknowledgements}
This research was in part supported by ``Praise: Practice and peRformance Analysis Inspiring Social Education'' a STREP project funded by the European Commission under the FP7 programme (grant number FP7-318770) and ``Social Media, Public Health, and the Urban Poor'' funded by the Goldsmiths, University of London Research and Enterprise Office. We thank Akaash Kolluri for assistance with formatting the manuscript and references for arXiv.

\bibliographystyle{unsrt}
\bibliography{Mendeley}

\end{document}